\documentclass[aps,prd,twocolumn,superscriptaddress,floatfix]{revtex4}
\usepackage{amssymb}
\usepackage{graphicx}
\expandafter\let\csname equation*\endcsname\relax
\expandafter\let\csname endequation*\endcsname\relax
\usepackage{amsmath}
\usepackage{amsbsy}
\usepackage{amsthm}
\usepackage{bbm}
\usepackage{bm}
\usepackage{epsfig}
\usepackage{dsfont}

\usepackage[colorlinks,linkcolor=red,citecolor=brown,urlcolor=blue]{hyperref}

\usepackage{soul}
\usepackage[usenames,dvipsnames]{xcolor}

\newcommand{\be}{\begin{equation}}
\newcommand{\ee}{\end{equation}}
\newcommand{\ben}{\begin{eqnarray}}
\newcommand{\een}{\end{eqnarray}}

\begin{document}

\title{Quantum estimation of the Schwarzschild space-time parameters of the Earth}
\date{\today}

\author{David Edward Bruschi}
\affiliation{Racah Institute of Physics and Quantum Information Science Centre, The Hebrew University of Jerusalem, Givat Ram, 91904 Jerusalem, Israel}
\affiliation{School of Electronic and Electrical Engineering, University of Leeds, Leeds LS2 9JT, United Kingdom}
\author{Animesh Datta}
\affiliation{Clarendon Laboratory, Department of Physics, University of Oxford, OX1 3PU Oxford, United Kingdom }
\author{Rupert Ursin}
\affiliation{Institute for Quantum Optics and Quantum Information, Austrian Academy of Sciences, Boltzmanngasse 3, 1090 Vienna, Austria}
\affiliation{Vienna Center for Quantum Science and Technology, Faculty of Physics, University of Vienna, Boltzmanngasse 5, A-1090 Vienna, Austria}
\author{Timothy C. Ralph}
\affiliation{School of Mathematics and Physics, University of Queensland, Brisbane, Queensland 4072, Australia}
\author{Ivette Fuentes}
\affiliation{School of Mathematical Sciences, University of Nottingham, University Park, Nottingham NG7 2RD, United Kingdom}

\begin{abstract}
We propose a quantum experiment to measure with high precision the Schwarzschild space-time parameters of the Earth. The scheme can also be applied to measure distances by taking into account the curvature of the Earth's space-time. As a wave-packet of (entangled) light is sent from the Earth to a satellite it is red-shifted and deformed due to the curvature of space-time. Measurements after the propagation enable the estimation of the space-time parameters.  We compare our results with the state of the art, which involves classical measurement methods, and discuss what developments are required in space-based quantum experiments to improve on the current measurement of the Schwarzschild radius of the Earth.
 \end{abstract}

\maketitle

\section{Introduction}

Quantum metrology uses nonclassical properties of the probes to estimate physical parameters with precisions higher than those attainable by classical methods, for equivalent resources. These techniques have been proposed to measure time, field strengths, and other parameters of interest~\cite{Giovannetti:LLoyd:11,Giovannetti2004}. The field began with the proposal of detecting gravitation waves with enhanced precision~\cite{quantumgravobs}, and in recent years, there has been a growing interest in applying quantum metrology techniques to estimate gravitational fields and accelerations~\cite{Sorrentino2011}. Measurement of physical quantities that play an important role in relativity are also of great relevance lie in practical applications. For example, a more precise knowledge of the Earth's space-time parameters has the potential of enhancing the performance of the Global Positioning System (GPS) since it requires relativistic corrections to determine time and position accurately.  Interestingly, almost all of these schemes employ non-relativistic quantum mechanics which is known to be incompatible with relativity.

A solution to this problem is to work within quantum field theory, which describes the behavior of quantum fields in space-time. It is a semiclassical theory in the sense that matter and radiation are quantized while the space-time is classical \cite{Birrell:Davies:82}. This theoretical framework enables one to incorporate relativistic effects at the low energy regimes that are being reached by cutting-edge quantum experiments.  Recently, methods for the application of quantum metrology to quantum field theory have been introduced in \cite{Ahmadi:Bruschi:14,Ahmadi:Bruschi:Sabin:14}.
Interestingly, relativistic effects such as particle creation, can be exploited to measure accelerations with an optimal measurement precision that is higher than the non-relativistic counterpart \cite{Ahmadi:Bruschi:Sabin:14}.  

Quantum technologies have made great progress in the past years and many implementations are reaching relativistic regimes. There are advanced plans to implement quantum communication protocols using satellites \cite{Scheidl:Wille:13,Ursin:etal:08} and to place quantum clocks in space~\cite{Pape:Terra:10}. However, the effects of gravity and motion on quantum properties and their applications have largely been ignored \cite{Alsing:Fuentes:12,Rideout:12}.  Relativistic quantum metrology and relativistic quantum information theory aim at developing quantum technologies taking into account relevant space-time effects. Within this theory, it has been shown that space-time can affect the implementation of quantum cryptography in satellite-based setups \cite{Bruschi:Ralph:14}. Furthermore, experiments involving Bose-Einstein Condensates (BECs) in space, have recently been proposed to test the effects of gravity on entanglement \cite{Bruschi:Sabin:14}. Relativistic quantum metrology has also been used in proposals to measure the Unruh temperature at accelerations within reach of current experiments \cite{Aspachs:Adesso:10},  to measure space-time parameters within algebraic quantum field theories \cite{Downes:Milburn:11} and to detect gravitational waves using BECs \cite{Sabin:Bruschi:14}.

In this paper, we apply relativistic quantum metrology \cite{Ahmadi:Bruschi:14,Ahmadi:Bruschi:Sabin:14} to estimate the Schwarzschild space-time parameters of the Earth and we propose an implementation using realisable space-based experiments. An entangled light wave-packet is sent from Earth to a satellite (or equivalently in the opposite direction). During its propagation, the wave-packet is red-shifted (blue-shifted in case of propagation in the opposite direction) and deformed due to the curvature of space-time. Measurements after the propagation enable the estimation of the space-time parameters. Currently, the measurement of the Schwarzschild radius has a relative error of $2\times10^{-9}$ \cite{Hirt:Cleassens:13}. The measurement method involved is classical in the sense that quantum estimation techniques are not involved.  We discuss what developments must take place in space-based quantum experiments in order to improve by an order of magnitude such classical measurements.

This work is organised as follows: in Sec.~\ref{tools} we will model the space-time of the Earth using the Schwarzschild metric for a spherical non-rotating mass and present the quantum field theory of a massless uncharged bosonic field describing the curvature effects on wave-packets that propagate from Earth to a satellite located at a fixed height.  In Sec.~\ref{relativistic:quantum:metrology} we will introduce relativistic quantum metrology techniques to estimate with high precision  parameters that appear in relativistic setups. In Sec.~\ref{estimation} we show how to apply these techniques, using single and two-mode quantum channels, to estimate lengths and the Schwarzschild radius of the Earth. In Sec.~\ref{estimation:of:errors} we compare our estimation method with the state of the art and comment on the use of multiple parameter estimation tools in this work. Finally, we present concluding remarks in Sec.~\ref{conclusions}.

\section{Light wave-packets propagating on Earth's space-time \label{tools}}
In this section we will present an approximate model for Earth's space-time and describe how wave packets of the electromagnetic field propagate in it (these techniques were introduced in \cite{Bruschi:Ralph:14}). Earth's space-time can be modeled by $(3+1)$-dimensional Schwarzschild metric which approximately describes the space-time outside a spherical non-rotating body \cite{Misner:Thorne:73,Wald:84}.  We are particularly interested in the propagation of wave-packets from a source on Earth to a receiver satellite (or in the opposite direction) situated at a fixed distance from the source. Assuming that Earth's angular momentum is negligible and considering that sources are small compared to the characteristic frequencies involved, we can consider only radial propagation. In this case, we can restrict our analysis to $1+1$-dimensions for which simple and analytical solutions of the Klein-Gordon equation are available. We employ Schwarzschild coordinates $x^{\mu}=(t,r)$, where $r$ is the proper distance from the singularity and  $t$ is the Schwarzschild proper time. This is assuming that  the observers are at spatial infinity  \cite{Misner:Thorne:73,Wald:84}. Important parameters in the analysis are Earth's mass $M$ and radius $r_E$.  The Schwarzschild metric $g_{\mu\nu}$ in the vacuum outside Earth is given by $g_{\mu\nu}=\text{diag}(-f(r),1/f(r))$, where $f(r):=1-\frac{r_s}{r}$, $r_S:=\frac{2GM}{c^2}$ is Earth's Schwarzschild radius and $G$ is the gravitational constant. The line element is given by $ds^2:=g_{\mu\nu}dx^{\mu}dx^{\nu}=-f(r)dt^2+1/f(r)dx^2$.

We are interested in describing the space-time at $r_E\leq r$.  Note that the Earth's radius is much larger than its Schwarzschild radius $r_S/ r_E\sim1.4\times10^{-9}$.
Free falling observers follow geodesics towards the centre of Earth. Satellites commonly employed for communication and GPS usually follow circular orbits of constant radius. These correspond to geodesics in the $3+1$ Schwarzschild metric for which acceleration is not required but require the observers to have angular momentum. In our analysis, we assume that observers Alice and Bob accelerate against the gravitational potential in order to remain at a constant distance from the Earth, i.e. at $r=r_A$ and $r=r_B$, respectively. A more detailed analysis of the effects of circular motion can be found in \cite{Bruschi:Ralph:14}.

An observer at constant distance $r=r_0$ must employ her own clock to measure time in her rest frame. The relationship between the proper time $\tau:=\frac{1}{c}\int ds$  of an observer at $r_0$ and the Schwarzschild time $t$ corresponding to the proper time of an observer at infinity is given by
\begin{eqnarray}
\tau=\sqrt{f(r_0)}\,t.\label{proper:time}
\end{eqnarray}
Now we describe the propagation of light from Earth to a satellite taking into curvature effects.

We consider an uncharged bosonic massless scalar field $\Phi(t,x)$ which is a good approximation to the longitudinal (or transverse) modes of the electromagnetic field \cite{Leonhardt:05,Friis:Lee:13}. The field $\Phi$ obeys the Klein-Gordon field equation $\square\Phi=0$, where the d'Alambertian  in curved space-time is given by $\square:=\frac{1}{\sqrt{-g}}\partial_{\mu}\sqrt{-g}\partial^{\mu}$ and $g:=\text{det}(g_{\mu\nu})$ \cite{Birrell:Davies:82}. The solutions to the equation are known as modes. Optical pulses propagating in the space-time can be modeled by wave-packets corresponding to an infinite superposition of mode solutions \cite{Leonhardt:05}.   

We solve the Klein-Gordon equation $\square\Phi=0$ in the $1+1$-dimensional Schwarzschild space-time using Eddington-Finkelstein advanced and retarded coordinates $u,v$ defined by $u:=c\,t-r_S$ and $v:=c\,t+r_S$, where the tortoise coordinate $r_S$ is defined as $r_S:=r+r_S\ln|\frac{r}{r_S}-1|$, see \cite{Misner:Thorne:73,Wald:84}. In these coordinates the Klein Gordon equation takes the form $\partial_u\partial_v\Phi(u,v)=0$ and the solutions are given by modes of the form $\phi^{(u)}_{\omega}(u)=\frac{e^{i\omega u}}{2\sqrt{\pi\omega}}$ and $\phi^{(v)}_{\omega}(v)=\frac{e^{i\omega v}}{2\sqrt{\pi\omega}}$, which represent outgoing and ingoing waves that follow geodesics $u=$const and $v=$const, respectively. The frequency $\omega>0$ is the frequency as measured by an observer infinitely far from Earth with respect to his proper time $t$. 
These modes are normalized through the inner product as $(\phi_{\omega}^{(u)},\phi_{\omega'}^{(u)})=(\phi_{\omega}^{(v)},\phi_{\omega'}^{(v)})=\delta(\omega-\omega')$, while mixed inner products vanish \cite{Birrell:Davies:82}. The solutions are eigenfunctions of the time-like Killing vector $i\partial_t$ and therefore satisfy the eigenvalue equation $i\partial_t\phi^{(u)}_{\omega}=\omega\phi^{(u)}_{\omega}$ and analogously for the $v$-modes.
Therefore, the time-like vector field $i\partial_t$ enables one to distinguish between positive $\phi_{\omega}^{(v)}$ and negative  $\phi_{\omega}^{(v)*}$ modes in a standard way. By associating creation/annihilation operators to positive/negative modes we expand the quantized  field operator $\Phi$ as
\begin{eqnarray}
\Phi=\int_0^{+\infty}d\omega\left[\phi_{\omega}^{(u)}a_{\omega}+\phi_{\omega}^{(v)}b_{\omega}+\text{h.c.}\right].
\end{eqnarray}
The bosonic annihilation operators obey commutation relations $[a_{\omega},a^{\dagger}_{\omega'}]=[b_{\omega},b^{\dagger}_{\omega'}]=\delta(\omega-\omega')$,
where mixed commutators vanish. The vacuum state $|0\rangle$ of the field is defined by the action of annihilation operators $a_{\omega},b_{\omega}$ as $a_{\omega}|0\rangle=b_{\omega}|0\rangle=0$.

Realistic photon sources do not produce monochromatic photons. A photon can be modeled by a wave packet: a distribution $F(\omega)\in\mathbb{C}$ of modes peaked around a frequency $\omega_0$ \cite{Downes:Ralph:13,Leonhardt:05}. The annihilation operator for the photon for an observer infinitely far from Earth, takes the form
\begin{eqnarray}
a_{\omega_0}(t)=\int_0^{+\infty}d\omega\, e^{-i\omega t} F_{\omega_0}(\omega)\,a_{\omega}\label{schwarzschild:wave:packet}.
\end{eqnarray}
The photon's creation and annihilation operators $a^{\dagger}_{\omega_0},a_{\omega_0}$ satisfy the canonical equal time bosonic commutation relations $[a_{\omega_0}(t),a^{\dagger}_{\omega_0}(t)]=1$ if the frequency distribution $F(\omega)$ is normalized, i.e. $\int_{\omega>0} d\omega |F(\omega)|^2=1$. This distribution naturally models a photon which is a wave packet of the electromagnetic field that propagates and is localized in space and time \cite{Bruschi:Lee:13}.

We consider that Alice, an observer on the surface of the Earth (i.e. $r_A=r_E$), sends a pulse to Bob who is on a satellite at constant radius $r=r_B>r_E$.
Alice and Bob measure frequencies in their laboratories \textit{with respect to their clocks}, i.e., with respect to their proper times $\tau_A$ and $\tau_B$. By employing the definition of proper time \eqref{proper:time} its easy to show that the eigenvalue equation for the modes takes the form $i\partial_{\tau_K}\phi^{(u)}_{\Omega_K}=\Omega_K\phi^{(u)}_{\Omega_K}$, where $K=A,B$ labels Alice or Bob and analogous formulas hold for $\phi^{(v)}$. This equation defines the physical frequency $\Omega_K$ measured by an observer at $r_K$ as $\Omega_K=\frac{\omega}{\sqrt{f(r_K)}}$.
Using the fact that  $\omega t$ is observer independent, one shows that if Alice prepares a sharp frequency mode $\Omega_A$, Bob will receive a mode with frequency
\begin{eqnarray}
\Omega_B=\sqrt{\frac{f(r_A)}{f(r_B)}}\Omega_A,\label{physical:frequency:relation}
\end{eqnarray}
This is the well-known gravitational redshift formula \cite{Misner:Thorne:73}. The equation implies that $\tau_B=\sqrt{\frac{f(r_B)}{f(r_A)}}\tau_A$. We are interested in finding how a wave-packet is transformed by its propagation from Alice to Bob on the Schwarzschild background. The wave packets  are given by 
\begin{eqnarray}
a_{\Omega_{K,0}}(\tau_K)=\int_0^{+\infty}d\Omega_K\, e^{-i\Omega_K \tau_K} F^{(K)}_{\Omega_{K,0}}(\Omega_K)\,a_{\Omega_K},\label{wave:packet}
\end{eqnarray}
where $K=A,B$ labels either Alice or Bob, $\Omega_K$ are the physical frequencies as measured in their labs using the proper times $\tau_K$ and $\Omega_{K,0}$ are the peak frequencies of the frequency distributions $F^{(K)}_{\Omega_{K,0}}$. We require that operators $a_{\Omega_K}$ satisfy the canonical commutation relations $[a_{\Omega_K},a^{\dagger}_{\Omega_K'}]=\delta(\Omega_K-\Omega^{'}_K)$. 
A wave packet $F^{(A)}_{\Omega_{A,0}}$ is prepared by Alice at time $\tau_A$. The wave-packet propagates radially and is received by Rob at a later time $\tau_B>\sqrt{f(r_B)/f(r_A)}\tau_A$. The wave packet received is modified due to curvature effects and is now given by $F^{(B)}_{\Omega_{B,0}}$. 
It is important to notice that Alice's and Bob's operators \eqref{wave:packet} can be used to describe the \textit{same} optical mode in two \textit{different} frames before and after propagation.

The relation between $a_{\Omega_A}$ and $a_{\Omega_B}$ was found in \cite{Bruschi:Ralph:14}, and can be used to find the relation between the frequency distributions $F^{(K)}_{\Omega_{K,0}}$ in different reference frames (or before and after propagation) \cite{Bruschi:Ralph:14}, 
\begin{eqnarray}
F^{(B)}_{\Omega_{B,0}}(\Omega_B)=\sqrt[4]{\frac{f(r_B)}{f(r_A)}}F^{(A)}_{\Omega_{A,0}}\left(\sqrt{\frac{f(r_B)}{f(r_A)}}\Omega_B\right).\label{wave:packet:relation}
\end{eqnarray}
We can see that the wave packet received by Bob has a different peak frequency and a different shape than the wave packet prepared by Alice. In particular, for the scenario of interest where Bob finds himself at higher altitudes than Alice ($r_B>r_A$), the wave-packet frequencies $\Omega_{B}$ measured by Bob will all be redshifted with respect to those created by Alice (see \eqref{physical:frequency:relation}).

We assume that Alice prepares a single photon in the mode $F^{(A)}_{\Omega_{A,0}}$ at time time $\tau_A=0$. The initial state is given by $|\psi_{\text{s.p.}}\rangle=a^{\dagger}_{\Omega_{A,0}}(0)|0\rangle$. Bob will receive the state $|\psi_{\text{s.p.}}\rangle=a^{\dagger}_{\Omega_{B,0}}(0)|0\rangle$, which is characterised by the distribution $F^{(B)}_{\Omega_{B,0}}(\Omega_B)$. 
Regardless of the specific model of the detector that Bob uses, if Bob uses a measuring device which is tuned to click when a photon in the wave packet $F^{(A)}_{\Omega_{A,0}}$ is received, the probability of the detector to click will be affected by the fact that he received a photon in the wave-packet $F^{(B)}_{\Omega_{B,0}}$ \cite{Bruschi:Ralph:14}.
Bob will conclude that the channel between him and Alice (i.e., the space-time) is noisy.  The quality of the channel can be quantified by employing the fidelity $\mathcal{F}$ defined as
\begin{eqnarray}
\mathcal{F}:=\text{Tr}^2\left[\sqrt{\sqrt{\rho}\rho^{\prime}\sqrt{\rho}}\right]\label{mixed:state:fidelity},
\end{eqnarray}
for arbitrary input states $\rho,\rho^{\prime}$. In case the input states are pure, for example  $\rho=|\psi\rangle\langle\psi|$ and $\rho^{\prime}=|\psi^{\prime}\rangle\langle\psi^{\prime}|$, the fidelity \eqref{mixed:state:fidelity} reduces to $\mathcal{F}=|\langle\psi|\psi^{\prime}\rangle|^2$ and the intensity fidelity gives the probability that the state was $\rho$ given that $\rho^{\prime}$ is obtained in a measurement. The fidelity $\mathcal{F}_{\text{s.p.}}$ of the channel in the case that Alice sends a single photon in a pure state is given by the overlap between the two distributions,
\begin{eqnarray}
\mathcal{F}_{\text{s.p.}}=\left|\Theta\right|^2\label{single:photon:fidelity},
\end{eqnarray}
where
\begin{eqnarray}
\Theta:=\int_0^{+\infty}d\Omega_B\,F^{(B)\star}_{\Omega_{B,0}}(\Omega_B)F^{(A)}_{\Omega_{A,0}}(\Omega_B).\label{single:photon:fidelity}
\end{eqnarray}
Clearly $\Theta=1$ for a perfect channel. If the curvature is strong enough, the distributions in \eqref{single:photon:fidelity} might have negligible overlap and the fidelity would be low. In the case of Earth-to-LEO communication, the fidelity is at least $75\%$, see \cite{Scheidl:Wille:13}.

The deformation of the wave packet due to the Earth's space-time curvature cannot be corrected by a linear shift of frequencies. While in \cite{Bruschi:Ralph:14} it was shown that the effects can have an impact on quantum communications in space, the main aim of this work is to take advantage of these effects to measure space-time parameters of interest, such as the Schwarzschild radius (mass of the Earth).  If Bob is aware that Alice is sending a wave packet $F^{(A)}_{\Omega_{A,0}}$, Bob can perform carefully selected measurements on his wave packet $F^{(B)}_{\Omega_{B,0}}$ and can estimate chosen parameters with great precision. 

It is always possible to decompose the mode $\bar{b}^{\prime}$ received by Bob in terms of the mode $b^{\prime}$ prepared by Alice  and an orthogonal mode $c^{\prime}$ (i.e. $[a^{\prime},c^{\prime\dagger}]=0$) \cite{Rohde:Mauerer:07}.
\begin{eqnarray}
\bar{b}^{\prime}=\sqrt{1-q}\,b^{\prime}+\sqrt{q}\,c^{\prime},\label{mode:decomposition}
\end{eqnarray}
where $q\leq1$.  The parameter $q$ is directly related to the fidelity of single photon transmission as defined in \eqref{single:photon:fidelity} by,\begin{eqnarray}
\Theta=\langle0\bigr|\bar{b}^{\prime}b^{\prime\dagger}\bigl|0\rangle=\sqrt{1-q},\nonumber
\end{eqnarray}
or equivalently $q={1-\Theta^2}$.

Lets assume that Alice and Bob employ real normalized Gaussian wave packets of the form
\begin{eqnarray}
F_{\Omega_0}(\Omega)=\frac{1}{\sqrt[4]{2\pi\sigma^2}}e^{-\frac{(\Omega-\Omega_0)^2}{4\sigma^2}}\label{Bob:wave:packet},
\end{eqnarray}
where $\sigma$ is the Gaussian width. In this case the overlap $\Theta$ is given by \eqref{single:photon:fidelity} where we extend the domain of integration to all the real axis. This is justified since the peak frequency is typically much larger than the spreading of the wave-packet ($\frac{\sigma}{\Omega_0}\ll1$), therefore, it is possible to include negative frequencies without affecting the value of $\Theta$. Using \eqref{Bob:wave:packet} and \eqref{wave:packet:relation} one finds that
\begin{eqnarray}
\Theta=\sqrt{\frac{2}{1+(1+\delta)^2}}\frac{1}{1+\delta}e^{-\frac{\delta^2\Omega_{B,0}^2}{4(1+(1+\delta)^2)\sigma^2}}\label{final:result},
\end{eqnarray}
where 
\begin{eqnarray}
\delta=\sqrt[4]{\frac{f(r_A)}{f(r_B)}}-1=-\frac{r_S}{4}\frac{L}{r_A(r_A+L)}.\label{delta:formula}
\end{eqnarray}
where $L=r_B-r_A$ the distance between Alice and Bob.
Notice that $\delta=0$ when Alice and Bob are in flat space-time ($f(r_A)=f(r_B)=1$) or when Alice and Bob are at the same height ($f(r_A)=f(r_B)$). Under these situations the modes perfectly overlap ($\Theta=1$) and, as expected, there are no effects due to gravity.

We are interested in the regime where $\delta\ll(\frac{\delta\Omega_{B,0}}{\sigma})^2\ll1$ (see \cite{Bruschi:Ralph:14}), which occurs for typical communication where $\Omega_{B,0}=700$THz and $\sigma=1$MHz. 
In this case, 
\begin{eqnarray}
		\Theta\sim1-\frac{\delta^2\Omega_{B,0}^2}{8\sigma^2},
\end{eqnarray}
In the following section we will introduce techniques of quantum metrology and their application to quantum field theory, where they can be employed to estimate with high precision relativistic parameters. 

\section{Relativistic Quantum Metrology\label{relativistic:quantum:metrology}}

In order to determine parameters that play a role in relativity such as proper time, proper accelerations and gravitational field strengths, it is necessary to work within the framework of quantum field theory in curved space-time \cite{Birrell:Davies:82}. This theory properly incorporates quantum and relativistic effects at regimes where space experiments take place. In this section we review general techniques and methods for the application of metrology to quantum field theory introduced in \cite{Ahmadi:Bruschi:14, Ahmadi:Bruschi:Sabin:14}.

In order to estimate the parameter with high precision it is necessary to distinguish two states $\Sigma_{\Theta}$ and $\Sigma_{\Theta+d\Theta}$ that differ by an infinitesimal change $d\Theta$ of the parameter $\Theta$. The operational measure that quantifies the distinguishability of these two states is the Fisher information \cite{Monras:Paris:07}.  Let us suppose that an experimenter performs $N$ independent measurements to obtain an unbiased estimator $\hat{\Theta}$ for the parameter $\Theta$. The Fisher Information $F(\Theta)$ gives a lower bound to the  mean-square error via the classical Cram$\mathrm{\acute{e}}$r-Rao inequality~\cite{Cramer:46} i.e., $\langle (\Delta \hat{\Theta})^{2}\rangle\geq\frac{1}{NF(\Theta)}$, where $F(\Theta)=\int\!d\lambda~p(\lambda|\Theta) (d\,\ln [p(\lambda|\Theta)]/d\lambda)^{2}$ and $p(\lambda|\Theta)$ is the likelihood function with respect to a chosen positive operator valued measurement (POVM)~$\{\hat{O}_{\lambda}\}$ with $\sum_{\lambda}\hat{O}_{\lambda}=\mathds{1}$. Optimizing over all the possible quantum measurements provides an even stronger lower bound~\cite{Braunstein:Caves:94} i.e.,
\begin{equation} \label{quantum:cramer:rao}
N\langle (\Delta \hat{\Theta})^{2}\rangle\geq\frac{1}{F(\Theta)}\geq \frac{1}{H(\Theta)},
\end{equation}
where $H(\Theta)$ is the quantum Fisher information. The optimal measurements for which the quantum Cram\'er-Rao bound \eqref{quantum:cramer:rao} becomes asymptotically tight can be computed \cite{Monras:Paris:07}. Unfortunately, these optimal measurements are usually not easy to implement in the laboratory. Nevertheless, in typical problems involving optimal implementations one can devise suboptimal strategies involving feasible measurements such as homodyne or heterodyne detection~\cite{Vidrighin:Donati:14}. In what follows we will present optimal metrology strategies based on the quantum Cram\'er-Rao bound and thus on the quantum Fisher information.

In relativistic quantum metrology one is interested in estimating parameters that are encoded in the evolution of states of a quantum field. It is therefore  convenient to work in the covariance matrix formalism \cite{Adesso:Illuminati:07}. This formalism enables elegant and simplified calculations for the application of quantum information and metrology techniques in relativistic quantum field theory~\cite{Ahmadi:Bruschi:14,Ahmadi:Bruschi:Sabin:14}. The formalism is applicable to bosonic fields as long as the analysis is restricted to Gaussian states. In this case, the states are described completely by their first moments $\langle X_{i}\rangle$ and their second moments, which are encoded in the covariance matrix $\Sigma_{ij}=\langle X_{i} X_{j}+X_{j}X_{i}\rangle-2\langle X_{i}\rangle\langle X_{j}\rangle$. Here $\langle\,.\,\rangle$ denotes the expectation value and the quadrature operators $X_{i}$ are the generalised position and momentum operators of the field modes. We follow the conventions used in~\cite{Ahmadi:Bruschi:Sabin:14,Friis:Fuentes:13}, where the operators for the $n$-th mode are given by $X_{2n-1}=\frac{1}{\sqrt{2}}(a_{n}+a^{\dag}_{n})$ and $X_{2n}=\frac{-i}{\sqrt{2}}(a_{n}-a^{\dag}_{n})$. 

Every unitary transformation in Hilbert space that is generated by a quadratic Hamiltonian can be represented as a symplectic matrix $S$ in phase space. These transformations form the real symplectic group $Sp(2n,\mathds{R})
$, the group of real $(2n\times2n)$ matrices that leave the symplectic form $\Omega$ invariant, i.e., $S\Omega S^{T}=\Omega$, where $\Omega=\bigoplus_{k=1}^{n}\Omega_k$ and $\Omega_k=-i\sigma_y$ and $\sigma_y$ is one of the Pauli matrices. The time evolution of the field, as well as the Bogoliubov transformations, can be encoded in this structure and are represented by a symplectic matrix. For example, the symplectic matrix corresponding to an arbitrary Bogoliubov transformation can be written in terms of the Bogoliubov coefficients $\alpha,\beta$ as
\begin{equation}\label{Bogosymplectic}
S=\left(
  \begin{array}{cccc}
    \mathcal{M}_{11} & \mathcal{M}_{12} & \mathcal{M}_{13} & \cdots \\
    \mathcal{M}_{21} & \mathcal{M}_{22} & \mathcal{M}_{23} & \cdots \\
    \mathcal{M}_{31} & \mathcal{M}_{32} & \mathcal{M}_{33} & \cdots \\
    \vdots & \vdots & \vdots & \ddots
  \end{array}
\right)\,,
\end{equation}
where the $\mathcal{M}_{mn}$ are the $2\times2$ matrices
\begin{equation}\label{Mmatrices}
\mathcal{M}_{mn}=\left(
                   \begin{array}{cc}
                     \Re(\alpha_{mn}-\beta_{mn}) & \Im(\alpha_{mn}+\beta_{mn}) \\
                     -\Im(\alpha_{mn}-\beta_{mn}) & \Re(\alpha_{mn}+\beta_{mn})
                   \end{array}
                 \right)\,.
\end{equation}
Here $\Re$ and $\Im$ denote the real and imaginary parts, respectively.  We consider a bosonic quantum field which undergoes a $\Theta$-dependent Bogoliubov transformation, where $\Theta$ is the parameter we want to estimate. For example, the transformation could be the expansion of the universe and the parameter in this case is the expansion rate \cite{Ball:Fuentes:06}. The covariance matrix after the transformation is given by $\Sigma_{\Theta}=S(\Theta)\,\Sigma_0 S^{T}\,(\Theta)$.  

In order to estimate $\Theta$ it is convenient to calculate the quantum Fisher information via its relationship to the fidelity $\mathcal{F}(\Sigma_{\Theta},\Sigma_{\Theta+d\Theta})$ which, in the density matrix formalism, is given by Eq. (\ref{mixed:state:fidelity}). The relation between the Fisher information $H(\Theta)$ and the fidelity $\mathcal{F}$ takes the form
\begin{equation}
 H(\Theta)=\lim_{d\Theta\rightarrow0}\frac{8\big(1-\sqrt{\mathcal{F}(\Sigma_{\Theta},\Sigma_{\Theta+d\Theta})}\big)}{d\Theta^{2}}. \label{quantumfishinfo}
\end{equation}
In this work we employ single Gaussian states and two-mode Gaussian states with vanishing first moments. In this case, the fidelity $\mathcal{F}(\Sigma_{\Theta},\Sigma_{\Theta+d\Theta})$ in the covariance matrix formalism is given by
\cite{Marian:Marian:12}
\begin{eqnarray}\label{one:mode:gaussian:fidelity2}
    \mathcal{F}(\Sigma_{\Theta},\Sigma_{\Theta+d\Theta})= 
    \frac{1}{\sqrt{\gamma}+\sqrt{\lambda}-\sqrt{(\sqrt{\gamma}+\sqrt{\lambda})^2-\sqrt{\eta}}},
\end{eqnarray}
where we have defined
\begin{align} \label{definition}
\gamma  &:=\frac{1}{16}\det(i\Omega\Sigma_{\Theta}i\Omega\Sigma_{\Theta+d\Theta}+\mathds{1})\,\\
\lambda &:= \frac{1}{16}\det(i\Omega\Sigma_{\Theta}\,+\mathds{1}) \det(i\Omega\Sigma_{\Theta+d\Theta}+\mathds{1})\,,\nonumber\\
\eta  &:=\frac{1}{16}\det(\Sigma_{\Theta}\,+\,\Sigma_{\Theta+d\Theta})
\end{align}
and $\mathds{1}$ is the identity matrix. The fidelity \eqref{one:mode:gaussian:fidelity2} for a single-mode Gaussian state with non-zero initial first moments is given by
\cite{Marian:Marian:12}
\begin{eqnarray}
    \mathcal{F}(\Sigma_{\Theta},\Sigma_{\Theta+d\Theta})\,=\, \label{one:mode:gaussian:fidelity}
    \frac{e^{\xi}}{\sqrt{\eta+\lambda}-\sqrt{\lambda}},
\end{eqnarray}
where $\xi=-(\langle\mathbb{X}\rangle_{\Theta+d\Theta}-\langle\mathbb{X}\rangle_{\Theta})^T\cdot(\Sigma_{\Theta}+\Sigma_{\Theta+d\Theta})^{-1}\cdot(\langle\mathbb{X}\rangle_{\Theta+d\Theta}-\langle\mathbb{X}\rangle_{\Theta})$. Notice that here $\lambda,\eta$ are defined as in \eqref{definition} except that the pre factor is $1/4$ instead of $1/16$ and the covariance matrices are $2\times2$ instead of $4\times4$.
We now employ these techniques to the particular case we are interested in analysing. 

\section{Estimating Earth's space-time parameters\label{estimation}}

In our work we are interested in estimating the Schwarzschild radius and propagation distance $L$ (between source and receiver) during which a wave-packet undergoes a transformation given by Eq. \eqref{mode:decomposition}. 
In the following we analyse the cases where Alice and Bob employ one-mode and two-mode entangled Gaussian states.

\subsection{Scheme employing a single-mode coherent channel}
In this scenario Alice and Bob exchange one mode systems (i.e., a single photon or a coherent state of a single mode). As discussed in the previous section, the mode $b'$ sent by Alice is received by Bob as a combination of the mode $b$ and the orthogonal mode $c$. We can therefore represent this process as a mixing (beam splitting ) of modes $b$ and $c$. The symplectic transformation of this pair of modes has a simple form since the Bogoliubov beta coefficients vanish,  the alpha coefficients are real with $\alpha_{bb}=\alpha_{cc}=\Theta$  and $\alpha_{bc}=\sqrt{(1-\Theta^2)}$. Therefore the symplectic matrix $S$ takes the form,
\begin{equation}\label{single:mode:bogoliubov:transformation}
S=\left(
\begin{array}{cc} 
\Theta \mathds{1}_2 &\sqrt{1-\Theta^2} \mathds{1}_2  \\ 
-\sqrt{1-\Theta^2} \mathds{1} _2 &\Theta \mathds{1} _2 
\end{array}\right) 
\end{equation} 
We assume Alice prepares her mode in a coherent state with parameter $\alpha\in\mathbb{C}$. The covariance matrix $\Sigma^{b,c}_0$ that represents the state is just the identity $\Sigma^{b,c}_0=\mathds{1}_{4\times4}$, while the first moments $\langle\mathbb{X}\rangle$ of the displaced state are
\begin{equation}
\langle\mathbb{X}\rangle=
  \begin{pmatrix}
    \alpha\\
    0 \\
    \bar{\alpha} \\
    0 
  \end{pmatrix}.
\end{equation}
The first moments $\langle\mathbb{X}\rangle_{\Theta}$ of the state \textit{after} beamplitting are
\begin{equation}
\langle\mathbb{X}\rangle_{\Theta}=
  \begin{pmatrix}
   \Theta\,\alpha\\
   -\sqrt{1-\Theta^2})\alpha \\
   \Theta\,\bar{\alpha} \\
   -\sqrt{1-\Theta^2})\bar{\alpha} 
  \end{pmatrix}.
\end{equation}
We wish to compute the QFI and for this we need the fidelity between the \textit{single mode} state after beam splitting and the same state with a small variation of the beam splitting parameter. We need to consider only the single mode quadratures $\langle\mathbb{X}_{\text{s.m.}}\rangle_{\Theta}$ defined as
\begin{equation}
\langle\mathbb{X}_{\text{s.m.}}\rangle_{\Theta}=
  \begin{pmatrix}
    \Theta\,\alpha\\
    \Theta\,\bar{\alpha}
  \end{pmatrix}.
\end{equation}
Finally, it turns out that the fidelity $\mathcal{F}_{\text{c.s.}}$ depends only on the first moments \eqref{one:mode:gaussian:fidelity} and has the expression
\begin{equation}
\mathcal{F}_{\text{c.s.}}=e^{-\frac{1}{2}(\Delta\mathbb{X}_{\text{s.m.}})^T\cdot\Delta\mathbb{X}_{\text{s.m.}}},
\end{equation}
where we have defined $\Delta\mathbb{X}_{\text{s.m.}}:=\langle\mathbb{X}_{\text{s.m.}}\rangle_{\Theta+d\Theta}-\langle\mathbb{X}_{\text{s.m.}}\rangle_{\Theta}$. We find $\mathcal{F}_{\text{c.s.}}=\exp[-|\alpha|^2\,d\Theta^2]$.
In our case we have that $\Theta=1-x$, where 
\begin{equation}\label{small:parameter:definition}
x:=\frac{\delta^2\Omega^2}{8\sigma^2}\ll1,
\end{equation}
where from now on we will omit the suffix in the expression of the frequency $\Omega_{B,0}$.
Therefore we find that the fidelity $\mathcal{F}_{\text{c.s.}}$ behaves to lowest order as
\begin{equation}
\mathcal{F}_{\text{c.s.}}=e^{-|\alpha|^2\,dx^2}.
\end{equation}
We can easily compute the relative error bound and we find
\begin{equation}
\frac{\Delta x}{x}\geq\frac{1}{2\sqrt{N}}\frac{1}{x\,|\alpha|}.\label{relative:bound:coherent:state}
\end{equation}

\subsection{Scheme employing a single-mode squeezed channel}
Here we consider that Alice prepares her mode $b$ initially in the most general single-mode state $\Sigma^{b,c}_0$. We also assume that the initial first moments vanish and therefore the numerator in \eqref{one:mode:gaussian:fidelity} trivially reduces to unity. 
We assume that mode $c$ is initially a thermal state for which the covariance matrix is the $2\times2$ matrix $\mu_b\mathds{1}_2$. Therefore, the total initial state is
\begin{equation}
\Sigma^{b,c}_0=
  \begin{pmatrix}
    B_-& B & 0 & 0\\
    B & B_+& 0 & 0 \\
    0 & 0 & \mu_b & 0 \\
    0 & 0 & 0 & \mu_b
  \end{pmatrix},
\end{equation}
where the coefficients of the matrix read $B_{\pm}=\mu_a \left(\cosh(2\,r)\pm\cos(2\,\psi)\sinh(2\,r)\right)$, $B=-\mu_a\sin(2\,\psi)\sinh (2\, r)$. Furthermore, the parameters $\mu_{a,b}:=\coth\bigl(\frac{\hbar\,\omega_{a,b}}{k_B\,T}\bigr)\geq1$ allow for initial extra mixedness, $\psi$ is an angle and $r$ is the squeezing. Notice that $\mu_a=\mu_b=1$ for the total state to be pure.

We now apply the Bogoliubov transformation of equation \eqref{single:mode:bogoliubov:transformation} to the state $\Sigma^{b,c}_0$ and trace over the unwanted mode. We are left with the reduced state $\Sigma^{b}$ of the form
\begin{equation}
\label{cmatrix}
\Sigma^{b}=
  \begin{pmatrix}
    C_- & C \\
    C &  C_+ 
  \end{pmatrix},
\end{equation}
where the coefficients of the matrix are $C_{\pm}:=(1-\Theta^2)\mu_b+\Theta^2\mu_a (\cosh(2\,s)\pm\cos(2\,\psi)\sinh(2\,s))$ and $C:=-\sin(2\,\psi)\Theta^2\mu_a\sinh (2\, s)$.

We can compute the final fidelity as a function of all the parameters. We find
\begin{widetext}
\begin{equation}
\label{general:mixed:fidelity}
\mathcal{F}=1-\frac{\mu_a^2+\mu_a^4+\mu_b^2+\mu_b\,\mu_a\,\left(\mu_a\,\mu_b\cosh(4\,s)-2\,(\mu_a^2+1)\cosh(2\,s)\right)}{\mu_a^4-1}dx^2,
\end{equation}
\end{widetext}
where we have defined $\Theta=1-x$, we have expanded around the small parameter $x$, which was defined in \eqref{small:parameter:definition}, and retained the lowest order contributions. 
Notice that one cannot employ formula \eqref{general:mixed:fidelity} to start from an initial squeezed vacuum state of modes $b,c$ for which $\mu_a=\mu_b=1$. This happens because of the behaviour of the perturbative expansions of the quantities defined in \eqref{definition}. In order to consider the case of initially pure states we can proceed as in appendix \ref{appendix} and we find
\begin{equation}
\frac{\Delta x}{x}\geq\frac{1}{2\sqrt{N}}\frac{1}{\sqrt{x}\,\sinh r}.\label{relative:bound:single:mode:squeezed:state}
\end{equation}
We can also estimate which is the bound on the distance $L$ between Alice and Bob or the Schwarzschild radius $r_S$. To do this we employ equation \eqref{delta:formula} and the chain rule as discussed at the end of appendix \ref{appendix}.  The bound on the relative error in the estimation of the Schwarzschild radius is given by 
\begin{equation}
\frac{\Delta r_S}{r_S}\geq\frac{2\sqrt2\, \sigma \, r_A\, (r_A+L)}{\sqrt{N}\, \Omega\, r_S\, L\, \sinh r},
\end{equation}
while the bound on the relative error on the distance  between Alice and Bob is given by
\begin{eqnarray}
\frac{\Delta L}{L}=\frac{2\sqrt2\, \sigma\, (r_A+L)^2}{\sqrt{N}\, \Omega\, r_S\, L\, \sinh r}
\end{eqnarray}
Notice that $\frac{\Delta L}{L}/\frac{\Delta r_S}{r_S}=1+\frac{L}{r_A}$ which is very close to unity in our scenario. 

\subsection{Scheme employing two-mode entangled channels}

We now allow Alice and Bob to employ two mode systems, which might be entangled. The scenario can now be modeled by four modes, where Alice prepares a two mode squeezed state between two wave packets of modes $b_1$ and $b_2$ centred around frequencies $\Omega_1$ and $\Omega_2$.  As in the single mode case, we consider that the orthogonal modes to $b_1$ and $b_2$, denoted $c_1$ and $c_2$ respectively, are initially in the vacuum state. The covariance matrix for the initial state is given by
\begin{equation}
\Sigma^{b_1b_2c_1c_2}_0=\left(
\begin{array}{cc} \tilde\sigma(r) &0 \\ 0  & \mathds{1}_4  
\end{array}\right) 
\end{equation} 
where $ \mathds{1}_4$ is the $4\times4$ identity matrix and $\tilde\sigma{(r)}$ is the covariance matrix of a two mode squeezed state
\begin{equation}
\tilde\sigma(r)=\left(
\begin{array}{cc} \cosh{(2r)} \mathds{1}_2&\sinh{(2r)}\sigma_x \\ \sinh(2r)\sigma_x &\cosh{(2r)} \mathds{1} _2 
\end{array}\right) 
\end{equation} 
Analogously to the single mode case, we model the effects of propagation from Alice to Bob on the modes $b_1,b_2$ by the transformations
\begin{eqnarray}
\bar{b_1}&=&\Theta_1\,b_1+\sqrt{{1-\Theta_1^2}}\,c_1,\\
\bar{b_2}&=&\Theta_2\,b_2+\sqrt{{1-\Theta_2^2}}c_2.\nonumber
\end{eqnarray}
Once more, these transformations are Bogoloiubov transformations and can be encoded in the symplectic matrix $S$ given by
\begin{equation}
S=\left(
\begin{array}{cccc} 
\Theta_1 \mathds{1}_2 &0&\sqrt{
1-\Theta_1^2} \mathds{1}_2&0  \\ 0&\Theta_2 \mathds{1} _2 &0&\sqrt{
1-\Theta_2^2} \mathds{1} _2\\\sqrt{1-\Theta_1^2} \mathds{1}_2 &0&-\Theta_1 \mathds{1}_2&0  \\ 0&\sqrt{
1-\Theta_2^2} \mathds{1} _2 &0&-\Theta_2 \mathds{1} _2 
\end{array}\right)\nonumber 
\end{equation} 
The final state represented by $\Sigma^{b_1b_2c_1c_2}$ after propagation is  $\Sigma^{b_1b_2c_1c_2}=S\,\Sigma_0^{b_1b_2c_1c_2}\,S^{T}$. We then proceed to trace over the ancilla modes $c_1,c_2$ and we obtain the covariance matrix $\Sigma^{b_1b_2}$ of the modes $b_1$ and $b_2$ after propagation,
\begin{equation}\Sigma^{b_1b_2}=\left(
\begin{array}{cc} 
(1+2\sinh^2r\,\Theta_1^2 )\mathds{1}_2 &\sinh{(2r)}\,\Theta_1,\,\Theta_2\,\sigma_z  \\ \sinh{(2r)}\,\Theta_1\,\Theta_2\,\sigma_z &(1+2\sinh^2r\,\Theta_2^2 )\,\mathds{1}_2
\end{array}\right) 
\end{equation} 
Since $\Theta_1\sim1-\frac{\delta^2\Omega_1^2}{8\sigma^2}$ and $\Theta_2\sim1-\frac{\delta^2\Omega_2^2}{8\sigma^2}$ we can write the final reduced covariance matrix $\Sigma^{b_1b_2c_1c_2}$ to second order in $\delta$ as,
\begin{equation}
\tilde\sigma(r)=\left(
\begin{array}{cc} 
\Sigma_{11}\,\mathds{1}_2&\Sigma_{12}\,\sigma_x \\ \Sigma_{12}\,\sigma_x& \Sigma_{11}\,\mathds{1}_2
\end{array}\right) 
\end{equation} 
where $\Sigma_{11}= 1+2\sinh^2{r}\,(1-\frac{\delta^2\Omega_1^2}{4\sigma^2})$ and $\Sigma_{12}=\sinh(2r)\,(1-\frac{\delta^2}{8\sigma^2}(\Omega_1^2+\Omega^2_2))$.
Using equation \eqref{one:mode:gaussian:fidelity2}, employing the Taylor series $\Sigma_{\delta+d\delta}=\Sigma_{\delta}+\dot\Sigma_{\delta}d\delta+\frac{1}{2}\ddot\Sigma_{\delta}(d\delta)^2$ and the chain rule for derivatives we find that
\begin{equation}
\frac{\Delta r_S}{r_S}\geq\frac{8\,\sigma\, r_A\, (r_A+L)}
{\sqrt{N(\Omega_1^2+\Omega_2^2)}\,r_S\,L\,\sinh{r}}
\end{equation}
and
\begin{equation}
\frac{\Delta L}{L}\geq\frac{8\,\sigma\, r_A^2}
{\sqrt{N(\Omega_1^2+\Omega_2^2)}\,r_S\,L\,\sinh{r}}.
\end{equation}
Notice that, now, $\frac{\Delta r_S}{r_S}/\frac{\Delta L}{L}=\frac{r_A+L}{r_A}$ which is also very close to unity in this scenario.

\section{Comparison between schemes, estimation of the errors and estimation of multiple parameters\label{estimation:of:errors}}

\subsection{Comparison between schemes}
In order to choose the best estimation scheme we need to be able to compare the different schemes proposed in this work and analyse their performance with respect to some given fixed resource \cite{Monras:Paris:07}.
The resource available is the number of photons $\bar{n}$ (or equivalently the total energy) of the input state, which can be easily computed for the scheme that employs a coherent state and the scheme that employs squeezed states. We have respectively $\bar{n}=\,|\alpha|^2$ and $\bar{n}=\,\sinh^2r$.
This allows us to express the final bounds \eqref{relative:bound:coherent:state} and \eqref{relative:bound:single:mode:squeezed:state} respectively as
\begin{align}\label{relative:error:bounds:with:constraint}
\left.\frac{\Delta x}{x}\right|_{\text{c.s}}\geq&\frac{1}{2\sqrt{N}}\frac{1}{x}\frac{1}{\sqrt{\bar{n}}}\nonumber\\
\left.\frac{\Delta x}{x}\right|_{\text{s.s}}\geq&\frac{1}{2\sqrt{N}}\frac{1}{\sqrt{x}}\frac{1}{\sqrt{\bar{n}}}.
\end{align}
Notice that the bounds \eqref{relative:error:bounds:with:constraint} both scale as the inverse of the total number of photons. This is in agreement with the general results found, for example, in \cite{Monras:Paris:07}.
The bounds in \eqref{relative:error:bounds:with:constraint} can be used to estimate, for example, the Schwarzschild radius $r_S$. A closer inspection shows that the relative error bound for coherent states scales as $1/r_S^2$ while the relative error bound for squeezed states scales as $1/r_S$. This implies that, for small Schwarzschild radius $r_S$, squeezed states perform better than classical states, in agreement with \cite{Monras:Paris:07}.

\subsection{Estimation of errors\label{estimation:of:errors:section}}
Here we compute the relative error on the measurement of the parameters of interest. 
We consider that the wave-packets will be peaked at frequencies $\Omega_1=\Omega_2=400$THz with widths of $\sigma=1$MHz. These parameters correspond to the current state of the art in space-based experiments \cite{Jofre:11}. Gbit exchange regimes of $N=10^{10}$ are achieved with the aim of implementing future QKD protocols \cite{Erareds:Walenta:10,Jouguet:Kunz-Jacques:13}. This means that we can realistically consider repeating the experiment $N=10^{10}$ times per second. We assume that the experiment takes place in one second in order to assure that Alice is radially aligned with Bob's satellite.  The distance we consider between Alice and Bob is $L=3.6\times10^{6}$m, therefore, $r_B=42.37\times10^{6}$m.  This radius corresponds to typical orbits for geostationary satellites.  It is conceivable that, given the rate of development of current quantum technologies, exchange of single photons at such distances will be achievable. Currently squeezing of $r =1.5$ has been achieved in cutting edge experiments \cite{PhysRevLett.104.251102}, therefore 
\begin{equation}
\frac{\sigma r_A^2}{\Omega\sinh{r}\sqrt{N}r_S\,L}\sim5.8\times10^{-7}.
\end{equation}
where we considered that the Schwarzschild radius is $r_S\sim10^{-2}$m. Employing single-mode channels result in relative errors of $\frac{\Delta r_S}{r_S}\sim\frac{\Delta L}{L}\geq2.4\times10^{-5}$, while two-mode entangled states result in $\frac{\Delta r_S}{r_S}\sim\frac{\Delta L}{L}\geq4.8\times10^{-5}$.  These estimates include errors in all parameters involved. This means that in computing the error in, for example $r_S$, we also included errors in $L$ and $r_E$. However, the errors in $L$ and $r_E$ turn out to be negligible. The relative error in the measurement of Schwarzschild radius in the current state of the art is $2\times10^{-9}$ \cite{note2}. It is conceivable that in five years squeezing of $r=6$ might be achieved. In this case, by being able to make $N=10^{16}$ measurements per second or by being able to integrate such  amount of measurements over a longer period of time,  it will be possible to improve the current state of the art by one order of magnitude using our quantum technique.

The error estimation presented here assumes that our scheme does not suffer from any source of loss or noise. However, in practical implementations of quantum metrology schemes, losses and noise can't be ignored \cite{Lee:Huver:09,Monras:Paris:07,PhysRevA.83.063836,PhysRevLett.113.060502,PhysRevLett.106.153603}. Quantum communication between ground users and space links has become recently an active area of research driven by proposals fro global communication networks \cite{Scheidl:Wille:13,Ursin:etal:08,Xin:11,note}. The dominant mechanism of loss over long distances is diffraction, e.g. transmitter and receivers aperture as well as the distance and the strength of the atmospheric turbulences between them \cite{Klein:74,FRIED:66}.  Furthermore, the exchange of single photons between ground and satellites is haunted by many sources of noise, such as all sorts of background lights and tight pointing requirements of the optical antennas used. Exciting recent studies have pioneered the ability to send and receive photons from ground stations to space when such sources of noise are present \cite{Bonato:Tomaello:09,Bacco:Canale:13,Yin:13,Vallone:Bacco:14}. We leave more detailed studies of such schemes in the presence of noise and loss to future investigations.

\subsection{Quantum estimation of multiple parameters}
In \ref{estimation:of:errors:section} we have assumed that the quantity that is estimated is the only source of uncertainty. Of course, in realistic schemes, two or more parameters might give rise to comparable errors and therefore it is natural to look for an extension of our work to multiple parameter estimation schemes.

It is known that the simultaneous quantum-limited estimation of multiple phases is more advantageous than estimating them individually~\cite{Humphreys2013}. The price for this advantage is  the preparation of specialised multi-mode correlated probe states. However, it turns out that the single- and two-mode squeezed states are not appropriate for this task. To demonstrate this, we calculate the quantum Fisher matrix for the estimation of the two parameters (Schwarzschild radius $r_S$ and separation between the two observers $L=r_B-r_A$ ). The multi-parameter quantum Cram\'er-Rao bound~\cite{Helstrom1974}
\be
\label{qcrb}
\mathrm{Cov}(r_S,L) \geq Q^{-1}
\ee
sets the limit of the attainable precision, provided by extending the covariance matrix formalism~\cite{Monras2013} to
\be
Q_{i\,j}=2 (\partial_i\bar{\Sigma})^{T}(\Sigma\otimes\Sigma-\Omega\otimes\Omega)(\partial_j\bar{\Sigma}),
\ee
where $\partial_i\bar{\Sigma}$ is the derivative of $\bar{\Sigma},$ the vectorised form of $\Sigma$ with respect to parameter $i.$ Given the form of the covariance matrix $\Sigma^b$ in Eqn.~(\ref{cmatrix})
\ben
Q_{L\,L} = \frac{r^2_A\, r^2_S}{L^2\,(L+r_A)^2} Q_{r_S\, r_S}, \nonumber \\
 Q_{r_S\, L} = Q_{L\, r_S} = \frac{r_A\, r_S}{L(L+r_A)} Q_{r_S\, r_S}.
\een
It is immediately clear that right hand side of Eqn.~(\ref{qcrb}) is singular, since $\mathrm{Det}[Q]=0.$ This leads to a trivial quantum Cram\'er-Rao bound, implying that the single-mode squeezed state is not suitable for estimating the two parameters simultaneously. 
\\
The same result is obtained for the two-mode squeezed state, implying that the two parameters cannot be estimated simultaneously using such a state. However, a different geometry should succeed where the squeezed states have failed. The role of a gravitational field on an optical field is to shift its frequency, the gravitational red-shift. In the present geometry, the ratio of the shifts at the location of the two observers is combined in to a single parameter (see \eqref{physical:frequency:relation}). This quantity takes the role of a mode overlap, which can itself be estimated using single- and two-mode squeezed states with higher precision. Rather than taking the ratio, it should be possible to estimate the frequency shifts individually simultaneously with a different geometry, with the allied advantage for multi-parameter quantum estimation. We leave this for future work.

\section{Conclusions\label{conclusions}}
In this work we have applied recently developed tools in relativistic quantum metrology \cite{Ahmadi:Bruschi:14} and techniques developed to study the effects of gravity on entanglement distribution protocols \cite{Bruschi:Ralph:14} to measure space-time parameters using traveling wave-packets in space. We have estimated the ultimate error bounds on the precision of measurements of parameters of interest in relativistic setups, such as distances between users of quantum communication protocols or the Schwarzschild radius of the Earth. A better estimation of this parameters can produce higher precisions in GPS and other quantum metrology applications. We find that with current levels of squeezing and channel capacities our scheme does not outperform current proposals for ultra precise measurements. However, we discuss which improvements in quantum technologies are necessary to improve the state of the art in the measurement of Schwarzschild radius of the Earth. Our work aims at providing the first steps in the estimation of space-time parameters using quantum metrology techniques. Furthermore, our results suggest that with the expected advances in technology foreseen in the next few years, the scheme proposed in this work will quickly provide dramatical improvements to current setups.  It is well known in quantum metrology that the use of highly entangled states, such as N00N-states, can reach the Heisenberg limit \cite{Giovannetti2004,Giovannetti:LLoyd:11,Su:Yang:14}. The use of such states in our scheme could in principle lead to precisions several order of magnitude beyond the current state of the art. 

\section*{Acknowledgments}
We would like to thank Paolo Villoresi, Carlos Sab\'in, Mehdi Ahmadi, Olivier Minster, Christian Schwarz, Eric Wille, Zoran Sodnik and Luigi Cacciapuoti for useful comments and discussions.
D. E. B. was supported by  the I-CORE Program of the Planning and Budgeting Committee and the Israel Science Foundation (grant No. 1937/12), as well as by the Israel Science Foundation personal grant No. 24/12. I. F. acknowledges support from EPSRC (CAF
Grant No. EP/G00496X/2). A. D. was supported by an EPSRC Fellowship (EP/K04057X/1). R.U. was supported by the FFG project Nr. 4299236 as well at from ESA contract Nr. 1-6889/11/NL/CBi. D. E. B. is grateful to the University of Nottingham for hospitality.

\appendix
\vspace{0.5cm}
\section{Derivation of single mode fidelity\label{appendix}}
We assume that mode $c$ is initially the vacuum state for which the covariance matrix is the $2\times2$ identity matrix $\mathds{1}_2$. Therefore, the total initial state is
\begin{equation}
\Sigma^{b,c}_0=\left(\begin{array}{cc} \sigma_0 & 0 \\ 0 & \mathds{1}_2  \end{array}\right) ,
\end{equation} 
where $\sigma_0=\text{diag}(e^r,e^{-r})$.
After propagation, the reduced covariance matrix for mode $b$  is given by $\Sigma^{b}_{\Theta}=\Theta^2\sigma_0 +\sqrt{1-\Theta^2}\mathds{1}_2$ (here mode $c$ has been traced out).

We compute the Taylor series of the matrix $\Sigma^{b}_{\Theta+d\Theta}$ as $\Sigma^{b}_{\Theta+d\Theta}=\Sigma^{b}_{\Theta}+\dot\Sigma^{b}_{\Theta}d\Theta+\frac{1}{2}\ddot\Sigma^{b}_{\Theta}(d\Theta)^2$.  By substituting this  expansion in Eq. (\ref{one:mode:gaussian:fidelity}) we obtain the fidelity to second order 
\begin{eqnarray}
   \mathcal{F}\,=1+\frac{(\dot{\det\Sigma^{b}})^2-2(\det\Sigma^{b}-1)\det(\dot{\Sigma}_b)}{2((\det\Sigma^{b})^2-1)}(d\Theta)^2,
   \nonumber
\end{eqnarray}
where the \textit{dot} denotes the derivatives with respect to the parameter $\Theta$.  Since $\dot\Sigma^{b}=2\Theta\dot\Theta(\sigma_0-1)$,
we find
\begin{eqnarray}
\det\Sigma^{b}&=&1+4\sinh^2r\Theta^2(1-\Theta^2)\nonumber\\
(\det\Sigma^{b}-1)\dot{\det\Sigma^{b}}
&=& 8^2\sinh^4r\Theta^2{\dot\Theta}^2(1+2\Theta^4-2\Theta^2)\nonumber\\
\det\dot{\Sigma^{b}}&=&-16\Theta^2\dot\Theta^2\sinh^2r\nonumber\end{eqnarray}
which finally allows us to find the Fisher information ${H}(\Theta)$ as
\begin{eqnarray}
{H}(\Theta)=\frac{8\dot\Theta^2(1-2\Theta^2+2\Theta^4)}{(1-\Theta^2)(\Theta^2(1-\Theta^2)+{(2\sinh^2r})^{-1})}\nonumber
\end{eqnarray}
We are now able to compute the bound on the variance ${\Delta}\Theta$. We find
\begin{eqnarray}
{\Delta}\Theta\geq\frac{1}{\sqrt{NH}}=\frac{\sqrt{(1-\Theta^2)(\Theta^2(1-\Theta^2)+{(2\sinh^2r})^{-1})}}{\dot\Theta\sqrt{2N}\sqrt{(1-2\Theta^2+2\Theta^4)}}\nonumber
\end{eqnarray}
Notice that the derivative term $\dot{\Theta}$ takes into account estimation of parameters contained in $\Theta$, for example as in \eqref{small:parameter:definition}.

\bibliography{Biblio}{}

\end{document}